\documentclass[aip,apl,reprint,superscriptaddress]{revtex4-1}
\usepackage{amsmath}
\usepackage{graphicx}
\usepackage{bm}
\usepackage{mathptmx}
\begin{document}
\title{Probing spin entanglement by gate-voltage-controlled interference of current correlation in quantum spin Hall insulators}
\author{Wei Chen}
\affiliation{National Laboratory of Solid State Microstructures and Department of Physics, Nanjing University, Nanjing 210093, China}
\affiliation{Department of Physics and Center of Theoretical and Computational Physics, The University of Hong Kong, Pokfulam Road, Hong Kong, China}
\author{Z. D. Wang}
\email{zwang@hku.hk}
\affiliation{Department of Physics and Center of Theoretical and Computational Physics, The University of Hong Kong, Pokfulam Road, Hong Kong, China}
\author{R. Shen}
\email{shen@nju.edu.cn}
\affiliation{National Laboratory of Solid State Microstructures and Department of Physics, Nanjing University, Nanjing 210093, China}
\author{D. Y. Xing}
\affiliation{National Laboratory of Solid State Microstructures and Department of Physics, Nanjing University, Nanjing 210093, China}
\begin{abstract}
We propose an entanglement detector composed of two quantum spin Hall insulators and a side gate deposited on one of the edge channels. For an \textit{ac} gate voltage, the differential noise contributed from the entangled electron pairs exhibits the nontrivial step structures, from which the spin entanglement concurrence can be easily obtained. The possible spin dephasing effects in the quantum spin Hall insulators are also included.
\end{abstract}
\maketitle

It is well known that the entanglement reflects a kind of nonlocal correlation\cite{Einstein,Bell} and plays an important role in quantum information and computation science.\cite{Nielsen} Recently, the creation and detection of electronic entanglement in solid state systems have attracted much interest, for the large-scale implementation of quantum information and computation schemes.\cite{Loss} The crossed Andreev reflection in mesoscopic s-wave superconductor systems has already been confirmed,\cite{Hofstetter} which is regarded as an effective proposal for the generation of spin entangled electrons in solids.\cite{Lesovik}

Though the idea to utilize entanglement in solids is elegant, a direct experimental evidence is still challenging. Several proposals for spin entanglement detection have been put forward, including the Bell inequality tests\cite{Kawabata,Chen} and the measurement of the shot noise in a beam splitter setup.\cite{Burkard} The former is based on the local hidden variable theories\cite{Bell} and the latter utilizes the relation between the spin entanglement and the antisymmetry of the electron wave functions.\cite{Burkard} Most recently, we have suggested another detection scheme by use of the quantum eraser effect\cite{Scully} and the complementarity principle\cite{Zeilinger}, where the spin entanglement concurrence is measured by the Aharonov-Bohm oscillation of the current correlation.\cite{Chen2}

On the other side, the two dimensional quantum spin Hall insulator (QSHI) has received much attention recently, as a topological matter.\cite{Kane} There are fully gapped bulk states and gapless helical edge states in QSHIs protected by the band topology.\cite{Kane} The helical electrons have their spins and moving directions bounded together, which provides an opportunity for the all-electrical control of spins.\cite{Chen}

In this paper, we propose a spin entanglement detector constructed by two QSHIs and a side gate on one of the edge channels. When entangled electrons are injected separately into different edge channels, the interference pattern of their current correlation contains the information of the entanglement and is controlled by the side gate in an all-electrical manner. More remarkably, under an \textit{ac} gate voltage, the differential noise exhibits a notable step structure for an easy observation and the spin entanglement concurrence can be drawn from the heights of those steps.

\begin{figure}
\centering
\includegraphics[width=0.45\textwidth]{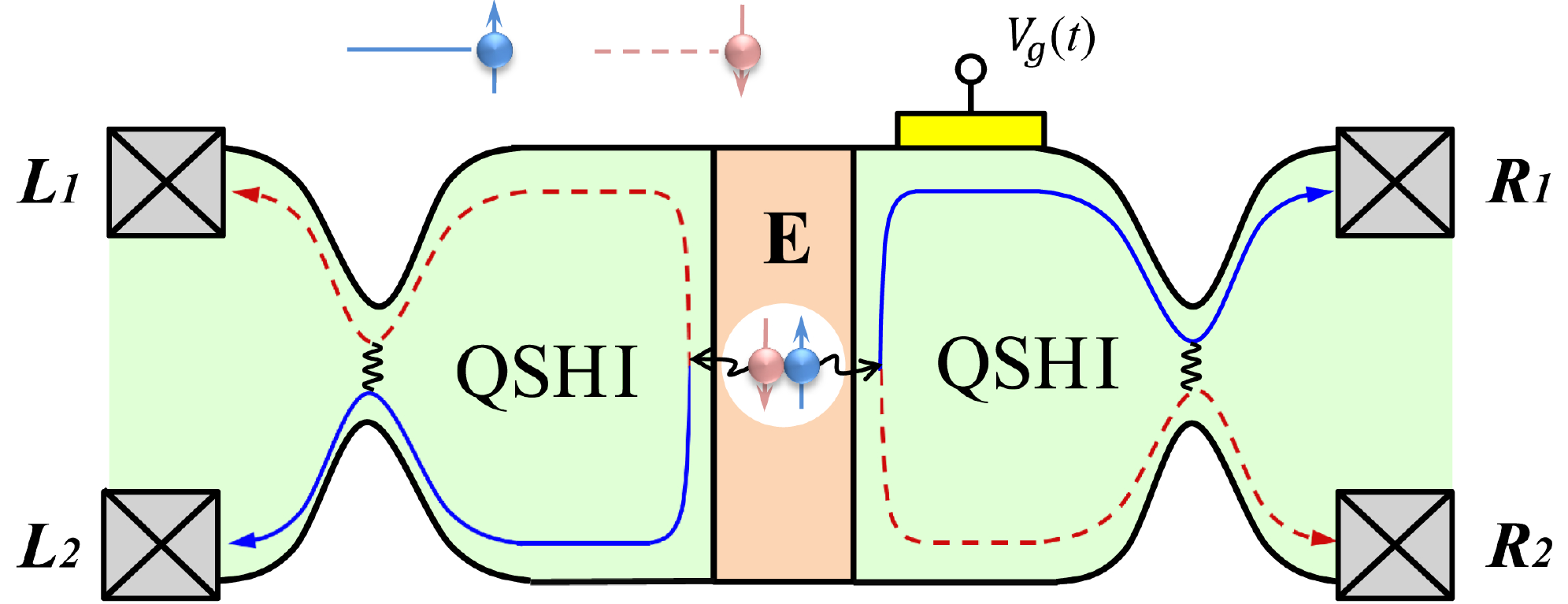}
\caption{(Color online) Illustration of the proposed setup. The entangled electrons are traveled separately from the entangler (labeled by E) into the helical edge channels of QSHIs. The spin up (down) channels are sketched as blue solid (red dashed) lines. The narrow regions in both sides represent the QPCs, which bring coupling between the upper and lower edge channels and serve as the beam splitters. A side gate is deposited on the upper channel in the right side. The currents and current correlations are measured in four terminals labeled by $L_{1,2}$ and $R_{1,2}$, respectively.}\label{fig1}
\end{figure}

The proposed setup is shown in Fig. \ref{fig1}, where the entangler in the middle region is weakly coupled to two QSHIs. When a bias voltage is applied between the entangler and the terminals, the entangled electrons will tunnel into the helical edge channels of QSHIs. We will investigate the current correlation between the left and right terminals to reveal the entanglement information. The entangled electrons may be injected into the same QSHI or separately into different QSHIs. However, the current correlation contributed from the former process is negligibly small in the weak coupling limit\cite{Chen,Samuelsson} so that only the latter process will be considered below. In each QSHI, there is a quantum point contact (QPC) depicted as the narrow region in Fig. \ref{fig1}, where the opposite edge channels are coupled together leading to the scattering between edges. It has been demonstrated that such a QPC can serve as an ideal beam splitter without any back scattering.\cite{Krueckl} A side gate is deposited on the upper edge of the right QSHI, which modulates the phase carried by the electron in that path.\cite{Dolcini}

The entangler can be realized by using superconductors\cite{Hofstetter,Lesovik} or quantum dots.\cite{Loss2} Here we adopt the former as an example, however, the proposed scheme is general and can be adapted to other kinds of entanglers. The superconductor is grounded and a bias voltage $-eV$ (less than the superconducting gap) is applied to all the terminals, so that the Cooper pairs are split into the helical edges in the QSHIs\cite{Chen3}. In the weak coupling limit, the tunneling coefficient is assumed as $\chi\ll1$. The entangled electron state is given by\cite{Chen,Samuelsson}
\begin{equation}\label{wave}
\begin{split}
\big|\Psi\big\rangle=\chi\int_{-eV}^{eV}d\varepsilon\big[&\sqrt{\kappa}a^{\dagger}_{L\uparrow}(\varepsilon)a^{\dagger}_{R\downarrow}(-\varepsilon)\\
&\mp\sqrt{1-\kappa}a^{\dagger}_{L\downarrow}(\varepsilon)a^{\dagger}_{R\uparrow}(-\varepsilon)\big]\big|0\big\rangle,
\end{split}
\end{equation}
where $|0\rangle$ represents the filled Fermi sea with a Fermi energy $-eV$, and $a^{\dagger}_{i,\sigma}(\varepsilon)$ creates an electron of energy $\varepsilon$ and spin $\sigma$ moving in side $i$, which satisfies the anticommutation relation $\{a_{i,\sigma}(\varepsilon),a^{\dagger}_{j,\sigma'}(\varepsilon')\}=\delta_{i,j}\delta_{\sigma,\sigma'}\delta(\varepsilon-\varepsilon')$. The helicity of the edge states indicates the spin-edge correspondence. For example, $a^{\dagger}_{L\uparrow}$ describes the creation of a spin-up electron moving left along the lower edge, as shown in Fig. \ref{fig1}. By changing the parameter $\kappa\in[0,1]$, the state in Eq. (\ref{wave}) can vary from the mostly entangled states ($\kappa=1/2$) to the product states ($\kappa=0, 1$). The measurement of the entanglement for a pure state can be described by the concurrence $C=2\sqrt{\kappa(1-\kappa)}$.\cite{Wootters} When $C=1$, the signs ``$\mp$'' in Eq. (\ref{wave}) correspond to the singlet and triplet entangled states, respectively.

After the transmission into the QSHIs, the electrons get scattered at the QPCs and then finally reach the terminals. The QPC is made of a narrow QSHI with coupling between two edges and the controllable spin-orbit coupling so that it can be regarded as an ideal beam splitter, where the electrons may change their edges during the forward scattering while the back scattering is completely ruled out.\cite{Krueckl} Thus, the operators for electrons in terminals $L_{1}$ and $R_{1}$ can be written as
\begin{equation}\label{operator}
\begin{split}
a_{L_{1}}=t_{2L}a_{L\uparrow}+t_{1L}a_{L\downarrow},\\
a_{R_{1}}=t_{1R}a_{R\uparrow}+t_{2R}a_{R\downarrow},
\end{split}
\end{equation}
where the coefficients $t_{1,2i}$ describe the amplitudes for the same-edge and the cross-edge transmissions in side $i$, respectively.

In order to probe the entanglement, we calculate the zero frequency noise power between terminals $L_{1}$ and $R_{1}$, which reads
\begin{equation}\label{noise}
S=2\int^\infty_{-\infty}dt[\langle\hat{I}_{L_{1}}(t)\hat{I}_{R_{1}}(0)\rangle-\langle\hat{I}_{L_{1}}(t)\rangle\langle\hat{I}_{R_{1}}(0)\rangle],
\end{equation}
where the current operator is defined as $\hat{I}_{i_{1}}(t)=(e/h)\times\int\int dEdE'e^{i(E'-E)t/\hbar}a^{\dagger}_{i_{1}}(E')a_{i_{1}}(E)$. The average in Eq. (\ref{noise}) is taken under the state Eq. (\ref{wave}), whose magnitude is of the order of $\chi$. Therefore, the leading term $\langle\hat{I}_{L_{1}}(t)\hat{I}_{R_{1}}(0)\rangle$ is of the order of $\chi^{2}$, while the term $\langle\hat{I}_{L_{1}}(t)\rangle\langle\hat{I}_{R_{1}}(0)\rangle$ is of the order of $\chi^{4}$ and can be neglected. By utilizing Eq. (\ref{operator}), we have
\begin{equation}\label{correlation}
\begin{split}
&\langle\hat{I}_{L_{1}}(t)\hat{I}_{R_{1}}(0)\rangle=\left(\frac{e\chi}{h}\right)^{2}\int^{eV}_{-eV}d\varepsilon \int^{eV}_{-eV}d\varepsilon 'e^{i(\varepsilon-\varepsilon ')t/\hbar}\times\\
&\left[\kappa T_{2L}T_{2R}+(1-\kappa )T_{1L}T_{1R}\mp C\sqrt{T_{2L}T_{2R}T_{1L}T_{1R}}\cos\varphi\right],
\end{split}
\end{equation}
where the transmission probabilities are $T_{1,2i}=|t_{1,2i}|^{2}$ and the total phase for electrons accumulated in a loop trajectory is $\varphi=\mathrm{Arg}(t_{2L}^{*}t_{1L}t_{2R}^{*}t_{1R})$. 

The cosine term in Eq. (\ref{correlation}) represents the interference effect. The right-moving electron can reach terminal $R_{1}$ by two paths. One is going in the upper edge first and then arriving in the terminal through the same-edge transmission at the QPC. The other is going in the lower edge first and then through the cross-edge transmission at the QPC. Due to the entanglement shown in Eq. (\ref{wave}) and the helicity of the edge states, the which-path information of the right-moving electron is registered by the spin of the left-moving one. As a result, the current in terminal $R_{1}$ would not show any interference behavior.\cite{Zeilinger} However, the current correlation between terminals $R_{1}$ and $L_{1}$ can still show interference. This is because that the left QPC can be regarded as a quantum eraser.\cite{Chen2,Scully} After scattering at the left QPC, the left-moving electrons arrive in terminal $L_{1}$ with mixed spins so that the which-path information is erased. The strength of the interference pattern of the current correlation reveals the information of the entanglement, as shown by the cosine term in Eq. (\ref{correlation}), which is proportional to the concurrence.  

Since there is a side gate on the upper channel in the right QSHI, phase $\varphi$ can be split into two parts $\varphi=\varphi_{g}(t)+\varphi_{0}$, with $\varphi_{g}(t)$ and $\varphi_{0}$ being the gate-dependent and the constant phase, respectively. Due to the helicity of the edge states, the gate voltage will not lead to any back scattering. For a slow-varying gate voltage, the period of the \textit{ac} gate is much longer than the traveling time of the electron in the gating region and the gate-dependent phase can be well approximated by $\varphi_{g}(t)=eV_{g}(t)d/\hbar v$, where $V_{g}$ is the gate voltage, $d$ is the length of the gating region, and $v$ is the Fermi velocity of the edge states. The constant phase is a parameter of the circuit and can be adjusted beforehand, e.g. by a \textit{dc} side gate. Here, we adopt $\varphi_{0}=0$ for simplicity. The \textit{ac} gate is assumed as a harmonic term $V_{g}(t)=V_{g}^{a}\sin\Omega t$ with the amplitude $V_{g}^{a}$ and the frequency $\Omega$. Phase $\varphi$ can be finally expressed by $\varphi =\varphi_{a}\sin\Omega t$ with $\varphi_{a}=eV_{g}^{a}d/\hbar v$.

Phase $\varphi$ is a function of period $2\pi /\Omega$ so that one can obtain the Fourier expansion 
\begin{equation}\label{Bessel}
e^{i\varphi}=\sum_{n=-\infty}^{+\infty}J_{n}(\varphi_{a})e^{in\Omega t},
\end{equation}
where $J_{n}(\varphi_{a})$ is the Bessel function of the first kind. Inserting Eq. (\ref{Bessel}) into Eqs. (\ref{noise}) and (\ref{correlation}), one obtains an elegant expression for the dimensionless differential noise (DN)
\begin{equation}\label{dnoise}
\begin{split}
\Lambda &=\frac{\partial S/\partial (eV)}{G_{0}}\\
&=1\mp C\big[J_{0}(\varphi_{a})+2\sum_{n=1}^{+\infty}J_{2n}(\varphi_{a})\Theta(eV-n\hbar\Omega)\big],
\end{split}
\end{equation}
where $\Theta$ is the Heaviside step function and the transmission probabilities are assumed as $T_{1,2L}=T_{1,2R}=1/2$ for the strongest interference. The conductance $G_{0}=\langle\hat{I}_{R_{1}}\rangle/V=e^{2}\chi^{2}/h$ can be measured by the current in terminal $R_{1}$, which is independent of the side gate.

\begin{figure}
\centering
\includegraphics[width=0.45\textwidth]{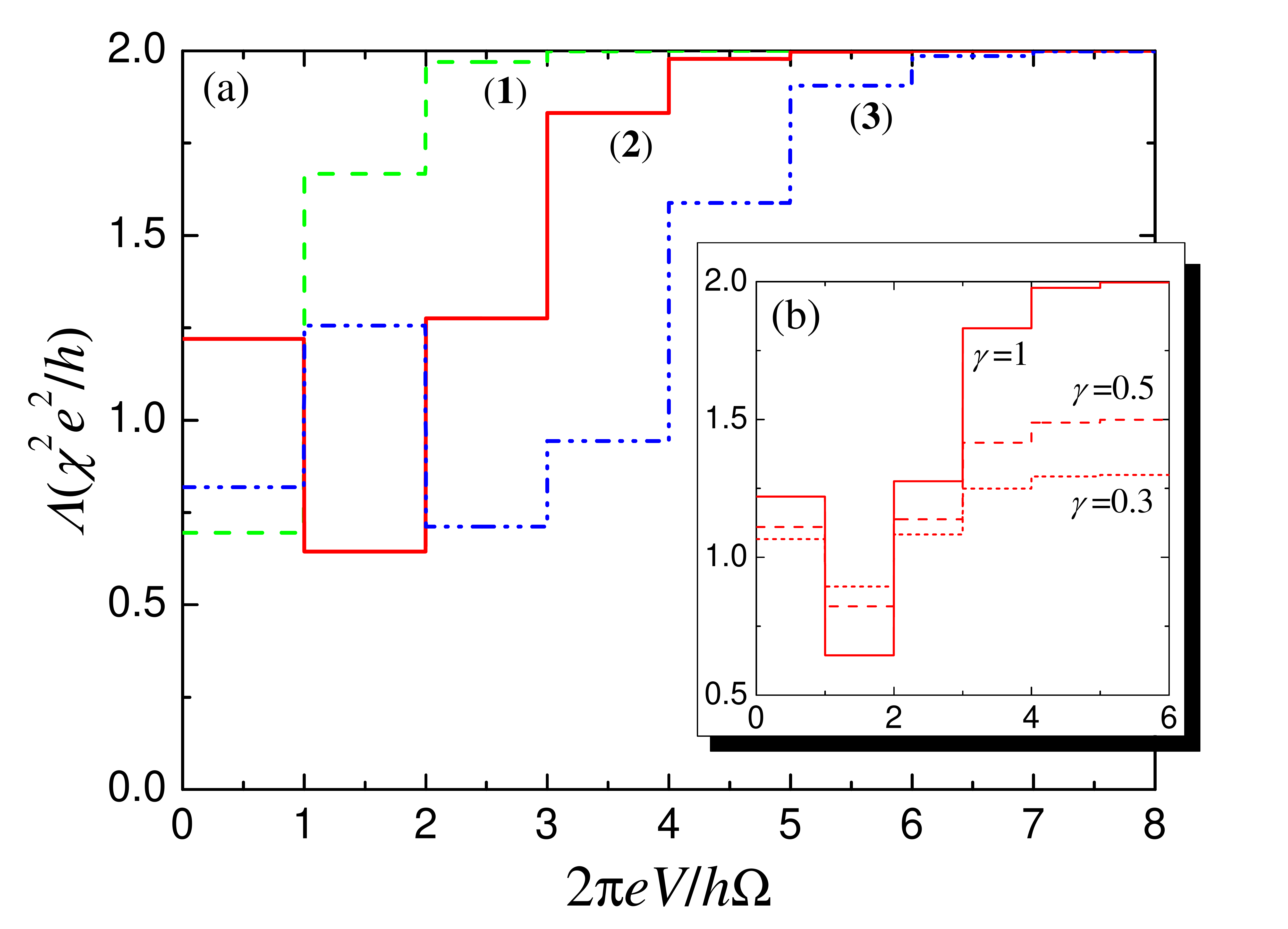}
\caption{(Color online) Plot of the DN as a function of the bias voltage for the triplet entangled states ($C=1$). (a) There is no spin dephasing and the phase amplitudes are $\pi$, $2\pi$, and $3\pi$ for the green (1), red (2), and blue (3) lines, respectively. (b) The phase amplitude is fixed at $\varphi_{a}=2\pi$ and the spin dephasing parameters are $\gamma=1, 0.5, 0.3$, respectively.} \label{fig2}
\end{figure}

A notable feature presented in Eq. (\ref{dnoise}) is the step structures of the DN. As long as the bias voltage increases by $\hbar\Omega/e$, the DN jumps onto a new step. The jump of the $n$th step is $\mp 2CJ_{2n}(\varphi_{a})$, where the sign $\mp$ here is the same as that in Eq. (\ref{wave}), depending on whether the entagled electron pairs are from the singlet or triplet superconductors. The step structure of the DN for the triplet entangled states of $C=1$ is plotted in Fig. \ref{fig2}(a) with various phase amplitudes. If the amplitude of the side gate and then the phase amplitude is known beforehand, the concurrence can be drawn from the height between any two adjacent steps. If the phase amplitude is not known, both the phase amplitude and the concurrence can be obtained by the heights of multiple steps. 

For a fixed phase amplitude, the Bessel function of a sufficiently high order approaches to zero rapidly. Therefore, as shown in Fig. \ref{fig2}(a), the jump between two steps becomes very small and the DN is saturated when the bias voltage is large enough. With the help of the identity $\sum_{n=-\infty}^{\infty}J_{n}=1$, one finds that the saturated value for the DN is $(1\mp C)$. The concurrence can be obtained by the saturated DN easily. We also note that the singlet and triplet entangled states can be distinguished by the saturated DN, which are less than and more than 1, respectively.  

Next, we discuss the spin dephasing effect in the QSHIs. Usually the phase accumulation in many-body environment satisfies Markov approximation, for a short memory time of the environment. By using the central limit theorem, the dephasing parameter is an exponential decay with the phase uncertainty, expressed as $\gamma=e^{-1/2\langle\delta\varphi^2\rangle}\in[0,1]$.\cite{Stern} With a finite spin dephasing, the entangled spin state Eq. (\ref{wave}) can be described by the density matrix $\hat{\rho}=\kappa|\uparrow\downarrow\rangle\langle\uparrow\downarrow|+(1-\kappa)|\downarrow\uparrow\rangle \langle\downarrow\uparrow|\mp\gamma\sqrt{\kappa(1-\kappa)}(|\uparrow\downarrow\rangle \langle\downarrow\uparrow|+|\downarrow\uparrow\rangle\langle\uparrow\downarrow|)$, and the average in Eq. (\ref{noise}) is calculated by $\mathrm{Tr}[\hat{\rho}\hat{I}(t)\hat{I}(0)]$. After some tedious algebra, one finds that Eq. (\ref{dnoise}) can be exactly recovered except that $C$ is replaced by $\gamma C$. The height of each step of the DN is suppressed by the spin dephasing in the QSHIs, as shown in Fig. \ref{fig2}(b). In principle, the $\gamma$ can be estimated from the spin decoherence length in the QSHIs. More practically, one can use an ideal entanglement source to calibrate the $\gamma$ in the proposed set up and then measure the concurrence of an arbitrary spin entangled state.

It is worthwhile to discuss the possible parameters for the realization of the proposed setup. The phase-gate relation employed in our model requires that the \textit{ac} gate varies slowly, \textit{i.e.}, the electron traveling time through the gating region being much smaller than the period of the gate. Taking the HgTe quantum wells as an example,\cite{Kane} the Fermi velocity of the edge states is given by $v\simeq5.5\times10^{5}$ m/s and the spin decoherence length is estimated as $1\mu\mathrm{m}$.\cite{Zhang} Given the scale of the proposed structure being $100\mathrm{nm}$, less than the spin decoherence length, the requirement for the frequency of the gate would be $\Omega\ll 30\mathrm{THz}$, which is always fulfilled in the usual case. In order to realize a phase amplitude of the magnitude of $\pi$, as shown in Fig. \ref{fig2}, the magnitude of the side gate is about $10\mathrm{mV}$. When the frequency of the gate is of the order of 1MHz and the bias voltage is of the order of 1nV, the magnitude of $eV/\hbar\Omega$ is of the order of 1 and a step structure of the DN similar to Fig. \ref{fig2} can be observed. With a smaller frequency and a larger bias voltage, one can obtain a much larger value for $eV/\hbar\Omega$. In this case, the step structure of the DN may be unclear, however, one achieves the saturated value for the DN directly, from which the spin entanglement concurrence is still obtained. 

In summary, we investigate the gate-voltage-controlled interference of the current correlation contributed by the spin entangled electron pairs in the QSHIs. The concurrence can be measured by the step structure of the DN. The helicity of the edge states in QSHIs makes the concurrence detection implemented in an all-electrical-controlled manner.

This work was supported by 973 Program (Grants No. 2011CB922100, No. 2011CBA00205, and No. 2009CB929504), by NSFC (Grants No. 11074111, No. 11174125, and No. 11023002), by PAPD of Jiangsu Higher Education Institutions, by NCET,  by the Fundamental Research Funds for the Central Universities, by the GRF (HKU7058/11P), CRF (HKU-8/11G) of the RGC of Hong Kong, and the URC fund of HKU. One of the authors (W. C.) would like to thank his friend Wei Xia for his tremendous support and encouragement.

\end{document}